\documentclass[lettersize,journal,onecolumn12]{IEEEtran}
\usepackage[dvips]{graphicx}
\usepackage{amsmath,amsfonts}
\usepackage{algorithmic}
\usepackage{algorithm}
\usepackage{array}
\usepackage[caption=false,font=normalsize,labelfont=sf,textfont=sf]{subfig}
\usepackage{textcomp}
\usepackage{stfloats}
\usepackage{url}
\usepackage{verbatim}
\usepackage{graphicx}
\usepackage{cite}
\usepackage{xcolor}
\usepackage{amssymb}
\usepackage{fontspec}
\begin{document}

\title{Noise Characterization and Robust Signal Detection in Yeast Pheromone Molecular Communication}

\author{Nikolaos~Ntetsikas,~\IEEEmembership{Student Member,~IEEE,} 
        Styliana~Kyriakoudi,
        Antonis~Kirmizis,
        Ioannis Krikidis,~\IEEEmembership{Fellow,~IEEE,}
        Ian F. Akyildiz,~\IEEEmembership{Life Fellow,~IEEE} and
        Marios Lestas,~\IEEEmembership{Member,~IEEE} 
\thanks{N. Ntetsikas and M. Lestas are with the Frederick Research Center and the Department of Electrical and Computer Engineering and Informatics, Frederick University, Nicosia, 1036, Cyprus (email: st023422@stud.frederick.ac.cy, eng.lm@frederick.ac.cy).}
\thanks{S. Kyriakoudi and A. Kirmizis are with the Laboratory of Epigenetics and Gene Regulation, Department of Biological Sciences, University of Cyprus, Nicosia, 2109, Cyprus (email: kyriakoudi.styliana@ucy.ac.cy, kirmizis.antonis@ucy.ac.cy).}
\thanks{I. Krikidis is with the IRIDA Centre and with the Department of Electrical and Computer Engineering,  University of Cyprus, Nicosia, 2112, Cyprus (email: krikidis.ioannis@ucy.ac.cy).}
\thanks{Ian F. Akyildiz is with Truva Inc., Alpharetta, Georgia, 30022, United States (email: ian@ianakyildiz.com).}}



\maketitle

\begin{abstract}

A critical aspect of Molecular Communications (MC) is the implementation of signal detection policies amidst noise. To date, noise characterizations within the MC field have predominantly drawn from methodologies found in wireless communications literature. In this study, we diverge from existing MC research by utilizing a newly developed experimental platform that employs yeast, allowing us to consider more realistic noise characterizations based on the relevant signaling pathways. We propose suitable signal detection mechanisms tailored to this experimental setup, which focuses on yeast cell-to-cell communications. Our analysis identifies gene transcription as the primary source of noise, and we utilize a Markov birth-death process model with Poisson arrivals and departures to characterize it. The noisy expression of the \textit{FUS1} gene is best represented using a mixed Gaussian distribution model. This model serves as a foundation for evaluating the performance of Maximum Likelihood Detection mechanisms in terms of Bit Error Rate (BER) for both symbol-by-symbol and sequence transmission schemes. Error analysis indicates that appropriate adjustments to the signal threshold can reduce errors to as low as 10\%, which is not negligible. In contrast, the detection of symbol sequences demonstrates enhanced error performance, achieving error rates as low as 0.4\%, albeit at the cost of increased computational complexity.

\end{abstract}





 \maketitle

\section{Introduction}\label{intro}

MC systems exhibit dynamic characteristics, and their performance is frequently influenced by unpredictable perturbation phenomena. At the microscale, the concentration of most chemical compounds forming reaction networks within confined environments, such as natural cells, is determined by chemical reactions among reactants. The behavior of these reactions significantly impacts gene expression and the abundance of proteins, proteases, enzymes, and other biomolecules. In the realm of mathematical modeling, various chemical interactions and events are often treated as deterministic; however, empirical studies have demonstrated that both molecular reaction and propagation phenomena are inherently stochastic \cite{bar2006noise}, \cite{kaern2005stochasticity}. Stochasticity within communication systems is commonly referred to as "noise", which can be identified across various communication sub-blocks, including the transmission, channel, and receiver (Rx) subsystems. Within the MC paradigm, each of these blocks incorporates distinct noise mechanisms that correspond to the different processes involved.

Modeling noise processes in bio-inspired MC systems is not always straightforward. To effectively incorporate noise models into the overall mathematical analysis, an understanding of the statistical or physical behavior of noise is essential. Most existing literature in the MC field \cite{Chang2018AdaptiveDA}-\cite{mosayebi2014receivers} adopts simplistic noise modeling approaches derived from wireless communications, such as additive Gaussian or basic Poisson counting models. While these modeling representations have proven adequate for approximating the binomial distribution of molecular counting events at the receiving end, intracellular processes such as transcriptional induction involve numerous molecular interactions, including the binding and release of transcription-associated factors to and from specific DNA elements over time. The unique nature of such processes necessitates a more nuanced consideration of noise mechanisms and models that can be integrated into microscale MC systems involving gene expression.

In this paper, we focus on two significant sources of noise based on the yeast pheromone testbed recently developed for investigating MC, as presented in \cite{10429943}. The first source pertains to free diffusion noise resulting from the Brownian motion of $\alpha$-factor particles as they traverse from the transmitting point to the receiver. The second source is transcriptional noise arising from the complexities associated with the generation of mRNA transcripts within the cell's nucleus following $\alpha$-factor stimulation in the form of impulse inputs. Our contributions can be summarized as follows:

\begin{itemize} \item We adopt appropriate noise models for both physical processes—namely, free molecular propagation and gene expression—and incorporate them into the mathematical models presented in \cite{10429943}, resulting in an overall stochastic channel model. \item For the first time, our yeast-pheromone system is utilized to compute the output of the mating gene \textit{FUS1} as a result of noisy gene expression, incorporating realistic noise mechanisms. \item Theoretical outcomes for the noisy expression of the \textit{FUS1} output gene are validated through GFP measurements on biological cells using our yeast experimental platform. \item We present two signal detection schemes to distinguish between the presence and absence of a received signal (single symbol case) in a noisy environment, as well as the detection of a sequence of symbols. Maximum Likelihood (ML) algorithms are employed for both schemes, revealing that sequence detection results in errors approximately two orders of magnitude lower than those associated with single symbol detection. Additionally, the basal (background) activation of \textit{FUS1} is considered as background noise in this analysis. \end{itemize}

The significance of the presented analysis and methodologies lies not only in the ability to validate theoretical findings using the available experimental testbed but also in the potential to leverage yeast as a foundation for developing biosensors that serve as MC receivers. This more accurate representation of extra- and intra-cellular signaling can bring MC closer to practical health applications.

The remainder of the paper is structured as follows: In Section \ref{Review}, we provide a comprehensive literature review on the state of the art regarding noise analysis and signal detection in MC systems. In Section \ref{Gene Noise}, we develop noise analysis tools tailored for microscale MC systems that incorporate cell-to-cell communication. Diverging from wireless literature, we adopt a modeling approach that aligns more closely with the biological nature of molecular interaction processes within a eukaryotic cell, such as the MATa haploid yeast. By integrating noise into our previously developed yeast communication model and validating theoretical findings with practical outcomes, we present a stochastic MC simulation tool that could benefit researchers engaged in microscale MC systems. In Section \ref{SigDet}, we outline signal detection policies for both symbol-by-symbol and symbol sequence reception, thereby establishing the initial framework for communications applicable to purely biological MC systems. Finally, in Section \ref{Conclusion}, we conclude this work with remarks and suggestions for future research directions.

\section{Literature review}\label{Review}

The complexities involved in the free diffusion of particles within isotropic microchannel liquid systems can lead to fluctuations in the number of particles detected by a receiver. These fluctuations, often referred to as particle counting noise or diffusion noise, have been the subject of investigation in various studies.

Additive diffusion noise models have been explored in several works, including those by \cite{Chang2018AdaptiveDA}-\cite{kuscu2016modeling}. In \cite{Chang2018AdaptiveDA},
 the authors employ the Normal distribution to characterize the concentration variations at the receiver unit, with a mean of zero and a variance dependent on chemical concentration. A similar data-driven approach is adopted by \cite{zhai2018anti}, where real particle measurements obtained from a spray transmitter and sensor receiver are utilized. They model the concentration fluctuations using a Normal distribution with a zero mean, adjusting the variance based on their experimental data. Additionally, \cite{shrivastava2020adaptive},  consider Poisson dynamics of an additive nature, calculating the mean value in accordance with the work of  \cite{ntouni2018optimal}. In the theoretical model presented by \cite{kuscu2016modeling}, two independent additive Gaussian noise components are incorporated to characterize the performance of bioFETs, with messenger molecules propagating within a microfluidic environment.

On the other hand, non-additive Poisson dynamics are employed to model MC channel noise in studies such as those  in \cite{mosayebi2014receivers}-\cite{pierobon2011diffusion}. In \cite{pierobon2011diffusion}, the authors analyze the free Brownian motion of individual molecules, which results in Poissonian dynamics regarding the total number of molecules arriving at the receiver unit's surface. This resulting noise, commonly referred to as "counting noise", represents fluctuations in molecular arrivals and is modeled in a non-additive manner. To account for the influence of Brownian motion of individual particles, the rate of the Poisson process is made dependent on the chemical concentration at each time point, rather than remaining fixed.

A similar methodology is presented in \cite{llatser2013detection}, where the authors utilize a non-additive Poisson statistical model to characterize the number of received molecules, 
following the framework established in \cite{pierobon2011diffusion}. 

In \cite{mosayebi2014receivers}, the authors also address Inter Symbol Interference (ISI), where the dynamic nature of the received molecular count is modeled as a Poisson random variable. Although the mean number of molecular arrivals per time interval is treated as constant, the authors account for noise contributions from previously transmitted molecules, which augment the fixed number of arrivals within a timeslot. 

This approach results in a more realistic model for propagation in free diffusion environments. We enhance the channel model presented in \cite{10429943} by adopting the modeling strategy outlined in \cite{Chang2018AdaptiveDA}. Our consideration of additive Gaussian noise is deemed an appropriate method for characterizing the noisy propagation of 
$\alpha$-factor molecules within the fluidic medium.

While diffusion noise is a fundamental aspect of MC systems in general, it is imperative to consider additional noise forms that significantly influence the temporal evolution of the received signal in bio-inspired MC systems. The inherent complexity of biological systems, particularly at the microscale, can introduce noise sources that dictate the overall fluctuations of the received signal at the receiving end. In cell-to-cell communications, the transmitting and receiving entities are represented by biological cells, which function as non-linear nanomachines incorporating numerous chemical reaction networks. These networks typically act as signal transducers, either facilitating molecular transmission or triggering gene expression in response to stimuli. Biological reaction networks are inherently dynamic and can introduce complexities to molecular-level chemical reactions at various stages of intracellular signal transduction. For example, in cells functioning as molecular receivers, noisy processes can influence the reactions of chemical species at multiple stages, from receptor activation to protein complex formation, and ultimately to transcription factor binding at DNA, culminating in gene expression. This intracellular noise can be categorized as "extrinsic" or "intrinsic", and is present in both prokaryotic and eukaryotic organisms \cite{kaern2005stochasticity}, \cite{swain2002intrinsic}. Distinguishing between intrinsic and extrinsic noise contributions is crucial. Intrinsic noise is associated with fluctuations inherent to gene expression, such as stochastic promoter activation/deactivation, variability in transcription factor binding, and mRNA and protein production and decay. In contrast, extrinsic noise arises from fluctuations in chemical rate constants and population variability of the chemical compounds involved in the biological cell's reaction network. This study primarily focuses on intrinsic noise.

Research on intrinsic and extrinsic noise has predominantly centered on prokaryotic organisms. In \cite{bicen2015efficient}, the authors investigate extrinsic noise in bacteria, employing a modified time-varying reaction rate equation to describe bacterial responses to molecular inputs. This noise mechanism correlates with fluctuations in protein molecules within bacterial populations, and the authors validate their findings through real experiments. The observed noisy behavior of bacteria is primarily attributed to the non-monotonic relationships between response delay, peak time, and response duration, alongside gene expression. Following identification, the theoretical formulation is grounded in Gaussian models. A similar investigation is presented in \cite{Bicen2020StatisticalMA}, which explores another extrinsic noise mechanism in bacteria: the rate and delay of bacterial responses to molecular stimulation. The authors characterize the probability distribution of response and delay noise sources and integrate these statistical measures into a single reaction rate equation to model bacterial output. In \cite{swain2002intrinsic}, the dynamics of both intrinsic and extrinsic noise sources in bacteria are studied, providing analytical models for each case. To model the noisy expression of mRNA and protein abundance, they propose a comprehensive yet straightforward reaction network, from which they derive analytical noise models, ultimately providing noisy outputs for mRNA and protein levels.

In contrast to prokaryotes, eukaryotic microorganisms exhibit greater complexity in the reaction networks involved in signal transduction pathways. Specifically, \textit{Saccharomyces cerevisiae} (\textit{S. cerevisiae}, hereafter) showcases numerous complex signal transduction pathways activated by various stimuli. The noise phenomena in this model organism are of particular interest as we seek to enhance our approach in \cite{10429943} by introducing stochasticity into our Receiver Model. Several studies have identified and modeled significant noise sources using either analytical or statistical frameworks to quantify the temporal dynamics of noisy gene expression. A comprehensive investigation into noise behavior in \textit{S. cerevisiae} is documented in \cite{bar2006noise}, where the authors conduct experimental measurements with live cells to quantify the relationship between noise strength and mRNA and protein abundance across 43 different genes expressed under 11 distinct environmental conditions. They propose an analytical model for the variance of gene expression and protein formation fluctuations, suggesting that mRNA and protein dynamics are governed by Poisson phenomena due to the stochastic birth and death of mRNA transcripts, as well as the randomness of gene "ON" and "OFF" states. Although a basic system of linear Ordinary Differential Equations (ODEs) is provided as a reference for gene expression, it is not integrated with the physical noise model, and no gene expression simulations were performed. In \cite{kaern2005stochasticity}, the authors examine the effects of promoter induction and transcriptional bursting fluctuations on mRNA and protein abundances in eukaryotes. They explore how variations in mRNA and protein levels influence fluctuation of noise strength, indicating that lower mRNA abundances can lead to noisier temporal dynamics. They also obtain experimental data from isolated gene measurements, concluding that intrinsic noise is a significant source of uncertainty in gene expression. Utilizing both simplistic deterministic models and stochastic simulations, they derive and juxtapose theoretical results for mRNA and protein concentrations. While the authors do not model the noise mechanism directly, they rely on stochastic simulation methods. A computational framework employing stochastic differential equations to simulate the time course of various transcriptional regulators in \textit{S. cerevisiae} cells is presented in \cite{10.1093/bioinformatics/bti415}. However, the focus of this study is on identifying potential regulators for various expressed genes, considering only additive Gaussian noise corresponding to the Brownian motion of regulatory protein molecules within the cell nucleus. A recent investigation into stochasticity in gene expression for \textit{S. cerevisiae} explores the significance of various intracellular noise sources \cite{Parab2020TranscriptionFB}. The authors utilize noisy gene expression data from previous studies to quantify metrics such as the Coefficient of Variation, which reflects noise levels in the temporal evolution of mRNA and protein abundance. They subsequently develop a statistical model, integrating multiple molecular features influencing gene expression, and compute metrics such as "R²" or "Fraction of variation explained" to identify the most relevant noise contributors. Notably, they find that the binding and unbinding of transcription factor molecules to the corresponding DNA promoters are the strongest predictors of noise in yeast. Furthermore, these binding/unbinding processes are best represented by Poisson dynamics. Similar to \cite{kaern2005stochasticity}, they employ stochastic simulations for gene expression while also incorporating simplified deterministic ODE models.

The aforementioned studies contribute to the understanding of noise dynamics in yeast and present theoretical results through stochastic simulation algorithms. However, they do not encompass realistic noisy dynamics across complete end-to-end signal transduction pathways in biological environments, neglecting the full spectrum of relevant biological processes and chemical reactions involved. By incorporating noise sources into the non-linear system of ODEs that comprise the Pheromone Response Pathway in \textit{S. cerevisiae} cells \cite{10429943}, we advance towards a more pragmatic representation of gene expression, such as that of the mating gene \textit{FUS1}, thereby enhancing the overall predictive capability of our model. Additionally, for the first time in the MC literature, we introduce signal detection policies for a biological system, exemplified by our previously developed yeast testbed. The algorithms we have developed focus on both symbol-by-symbol detection and symbol sequence detection within a noisy environment, considering the presence of channel memory.

\section{Gene Expression Under Noise} 

\label{Gene Noise}
The system under investigation is centered on the yeast mating pheromone response pathway. The budding yeast, \textit{S. cerevisiae}, exists in one of two mating types: MAT$\alpha$ and MATa. A communication circuit is established between these two types, as they secrete distinct peptide pheromones to identify nearby mating partners of the opposite type. Specifically, MATa cells release the a-factor pheromone, which is a 12-amino-acid-long peptide, while MAT$\alpha$ cells secrete the $\alpha$-factor pheromone, a 13-residue peptide. The pheromones released are detected by surface receptors on yeast cells of the opposing mating type \cite{2005ErratumT}.

In this study, we utilize the yeast pheromone testbed that we recently developed and presented in \cite{10429943} as the foundation for our investigation. The testbed comprises MAT$\alpha$ cells acting as senders and MATa cells as receivers, with $\alpha$-factor molecules functioning as the message carriers. Upon binding of the $\alpha$-factor to the receptors on the receiving cells, signaling pathways are activated, leading to the expression of the mating gene \textit{FUS1} as the system's output. A computational channel model was also introduced in \cite{10429943}, which was validated using experimental data. This mathematical model is represented by coupled nonlinear ODEs, and the objective of this work is to enhance the model by incorporating terms that account for the most significant sources of noise. This enhancement will facilitate a bit error analysis of appropriate signal detection mechanisms.

To model the stochasticity in \textit{FUS1}, we begin by integrating a Normal noise model into the concentration solution of $\alpha$-factor reported in \cite{10429943}. This noise model aligns with the findings presented in \cite{Chang2018AdaptiveDA}:

\begin{equation} \label{alpha_noisy}
    \alpha = \alpha_{CIR} + \mathcal{N}(0,\frac{\alpha_{CIR}}{V})
\end{equation}
Where V is the volume of the Rx (MATa) cell, and $\alpha_{CIR}$ represents the closed form for the impulse response of the fluidic channel. Here, $V=113$ $\mu m^3$ \cite{Milo2009BioNumberstheDO}.

In contrast to the simplistic additive noise observed in physical systems, transcriptional noise, characterized by Poisson binding and unbinding events, cannot be addressed in the same manner. The inherent nature of these biological processes is analogous to birth (binding) and death (unbinding) processes, which exhibit the Memoryless property \cite{lestas2008noise}. Our focus is on equation (58) from \cite{10429943}, which serves as a deterministic ODE that calculates the deterministic time course of the transcription regulator Ste12. For enhanced clarity, we present the equation that quantifies the time dynamics of Ste12, alongside the corresponding equation for the output gene \textit{FUS1}, derived from our previous research.

\begin{flalign}
&\frac{d[Ste12]}{dt} = v39-v40 \label{Ste12} &\\
&\frac{d[Fus1_{mRNA}]}{dt} = P_3k_{trFUS1} - d_{mRNA}[Fus1_{mRNA}] \label{FUS1 gene}&\\
&P_3 = \frac{[S2]}{[S2]+[Ste12]+[SD1]+[KD3]}
\end{flalign}

For further details regarding the ODEs that describe the yeast system, the reader is directed to \cite{10429943}. Thus far, our model presupposes that all Ste12 present within the cell will bind to and unbind from the promoters of \textit{FUS1} at a rate that is significantly faster than the changes in Ste12 abundance over time \cite{houser2012positive}. To incorporate Poisson dynamics, we develop a new differential equation that serves a dual purpose: to accurately depict the binding and unbinding processes of Ste12 molecules while simultaneously introducing stochasticity of a Poisson nature. Consequently, we introduce the following equation to our system:

\begin{equation}\label{Ste12_Poisson}
Ste12_{bound} = Poiss(\lambda_{in}\cdot t_s)-Poiss(\lambda_{out}\cdot t_s)
\end{equation}

In equation \eqref{Ste12_Poisson}, the parameters $\lambda_{in}$ and $\lambda_{out}$ denote the rates at which molecules bind to and unbind from the DNA promoter, respectively, over a specified time interval. This kinetic mechanism can be effectively modeled as an M/M/1 queue, with the DNA promoter acting as the server and the binding and unbinding events representing Poisson arrivals and departures. The expressions for $\lambda_{in}$ and $\lambda_{out}$ are provided below, where the concentration of Ste12 is indicated within brackets:

\begin{flalign}
&\lambda_{in} = [Ste12]\cdot k_{bind}&\label{lambdaIn}\\
&\lambda_{out} = [Ste12_{bound}]\cdot k_{unbind}\label{lambdaOut}&
\end{flalign}

The parameters  $k_{bind}$ and $k _{unbind}$ represent the binding and unbinding constant rates, respectively, and are measured in minutes$^{-1}$. The time period $t_s$, set at $1.5$ minutes, is a reasonable choice given the relatively slow dynamics of the system's overall response and the total duration of the experiment, which is $360$ seconds. Selecting such a brief sampling period allows us to model finer noise dynamics effectively.

The terms $\lambda_{in}$ and $\lambda_{out}$ are expressed in molecules per minute after appropriate adjustments. Consequently, we must convert the two feedback loops in equation \eqref{Ste12_Poisson} back to the chemical concentration scale to maintain consistency with the mathematical models established in our previous work. In this study, we have chosen $k_{bind}=7$ minutes$^{-1}$ and $k_{unbind}=3$  minutes$^{-1}$. Specific values for Ste12 transcriptional activity are not explicitly reported in the literature. However, in \cite{Das2017EffectOT}  experimental studies in mammalian systems demonstrated that many transcriptional factors bind to DNA transiently, with residence times ranging between seconds and minutes. Furthermore, a large number of yeast proteins have been reported to exhibit a maximum lifetime of 10 minutes. The residence time of specific transcriptional factors on promoters has been inferred to be of that order. Therefore, the selected values for $k_{bind}$ and $k_{unbind}$ in this work are drawn from the referenced range, which is considered reasonable for chemical reaction rates of this type.

\begin{figure}[!hbt]
\centering
\includegraphics[width=3in]{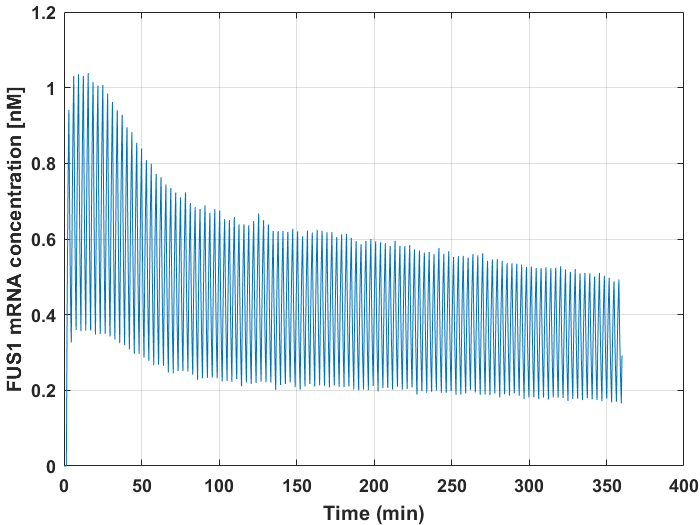}
\caption{Noise in \textit{FUS1} expression under stochastic transcriptional binding/unbinding.}
\label{noisy Fus1}
\end{figure}

Utilizing equations \eqref{lambdaIn} and \eqref{lambdaOut}, we quantify the time evolution of the gene \textit{FUS1} based on Poissonian bindings and unbindings, as illustrated in Figure \ref{noisy Fus1}, in response to a 10 $\mu$M stimulation of $\alpha$-factor. The computation of \textit{FUS1} expression levels over time is derived from equation \eqref{FUS1 gene}. The rapid oscillations observed in the response indicate that the stochastic nature of transcription factor binding leads to a "jittery" gene expression pattern, deviating from a smooth deterministic behavior.

\subsection{Experimental Platform}

\begin{figure}[!hbt]
\centering
\includegraphics[width=\columnwidth]{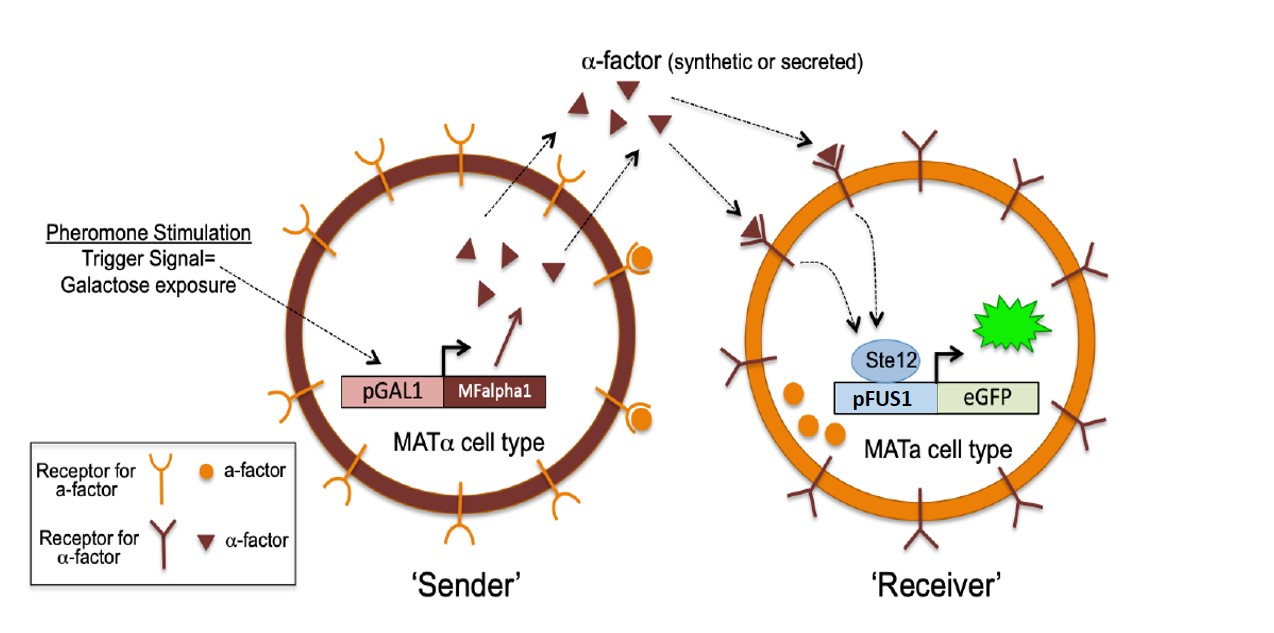}
\caption{Fundamental point-to-point yeast communication system \cite{10429943}.}
\label{System Description}
\end{figure}

The experimental yeast testbed utilized to obtain practical measurements of gene expression in MAT$\alpha$ cells is described in detail in \cite{10429943} and will be summarized here. The design of the testbed is based on the extensively studied biological process of the mating response in \textit{S. cerevisiae}, which employs MAT$\alpha$ cells as biological pheromone transmitters and MATa cells as biological receivers. The MATa cells indicate pheromone detection by emitting green fluorescence, which is captured by our testbed equipment (see Figure \ref{System Description}).

In preparation for conjugation, the pheromonal signals received by the yeast activate G-protein-coupled receptors on the MATa cells, leading to the activation of the mating gene-specific transcription factor Ste12. Ste12 has been demonstrated to induce the expression of the mating gene \textit{FUS1}, among others. Building on our previous work in \cite{10429943}, we have engineered MATa cells to express a reporter protein known as superfolder Green Fluorescence Protein (sfGFP) under the intrinsic \textit{FUS1} promoter, achieved through the simultaneous deletion of the endogenous \textit{FUS1} gene.

\subsection{Validation}

To evaluate the validity of our new model, we aim to compare our theoretical findings with experimental measurements, following the methodology outlined in \cite{10429943}. We conducted multiple experiments and computations to obtain both analytical and practical quantifications, selecting average values for validation purposes.

\begin{figure}[!hbt]
\centering
\includegraphics[width=3in]{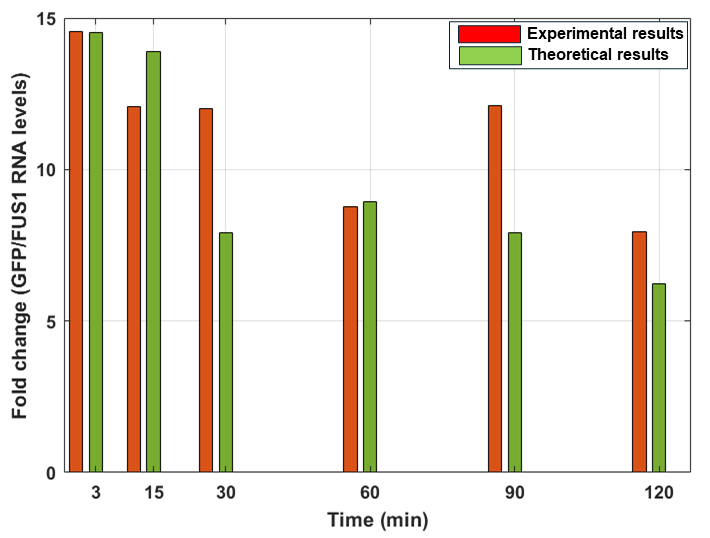}
\caption{Validation of theoretical and experimental results. Noise is included in the Receiver Model.}
\label{noisy validation}
\end{figure}

In this context, we quantify the fold change of the mRNA transcripts predicted by our model, using the basal (background) transcription "leakage" as the baseline. As illustrated in Figure\ref{noisy Fus1}, the fluctuations in RNA levels around the mean value indicate that the time evolution of \textit{FUS1} aligns more closely with the measured values of GFP RNA levels, as shown in Figure \ref{noisy validation}. However, some discrepancies remain, primarily related to amplitude mismatches.

The differences observed can be attributed to unaccounted biological processes inherent in actual transcription events, as well as variations between equipment functionality and simplistic mathematical models. Nevertheless, incorporating transcriptional noise into our computations yields a more diverse and nuanced response of the Rx overall.


\section{Signal Detection} \label{SigDet}

In this section, we propose a theoretical framework for detecting the upregulation of \textit{FUS1} following stimulation with $\alpha$-factor. Identifying such signals in realistic scenarios can be challenging due to the inherent noise in gene expression processes and the background expression of \textit{FUS1} in the absence of the stimulating factor. In this study, we develop detection policies for two distinct cases.

First, we examine the reception of a single symbol, distinguishing between two possibilities: either a pheromone impulse was transmitted (denoted as "1") or no transmission occurred (denoted as "0"). The simplicity of this binary transmission scheme allows us to employ the Maximum Likelihood Detection (MLD) method. This approach is particularly effective for straightforward scenarios, as it can accommodate arbitrary data and often minimizes the probability of erroneous reception \cite{kay1998fundamentals}.

Second, we address the reception of a series of symbols, where complex noise phenomena, along with inter-symbol interference (ISI) from adjacent symbols, render symbol-by-symbol detection impractical. Consequently, we develop a computational framework for detecting a sequence of transmitted symbols.

We derive the empirical distribution of the received amplitude (i.e., \textit{FUS1} output concentration) for both transmission and non-transmission scenarios, alongside probabilistic models that best fit these distributions. Distribution fitting was conducted on simulation datasets using MATLAB. The selection of the optimal fit was based on comparisons of the Akaike Information Criterion (AIC) across various distributions. The distribution with the lowest AIC value corresponds to a Gaussian Mixture Model (GMM) with three components, yielding an AIC of $-1.45 \cdot 10^4$. This distribution model can be characterized by the following Probability Density Function (PDF):

\begin{equation} \label{GMM}
    g_j(x|\mu_j, \sigma^2_j) = \sum_{i=1}^{n} \pi_{ji} f(x|\mu_{ji}, \sigma^2_{ji}) , \quad j=[0,1],\quad i \in \mathbb{Z+}
\end{equation}
and,
\begin{equation} \label{proportions}
     \sum_{i=1}^{n} \pi_{ji} = 1
\end{equation}

The function $f(x|\mu_{ji}, \sigma^2_{ji})$, represents the probability density function (PDF) of a single Gaussian distribution. The constant $\pi_i$ denotes the proportion of each Gaussian component included in the Gaussian Mixture Model (GMM). The mean and variance of each Gaussian component are represented by $\mu_{ji}$ and $\sigma^2_{ji}$, respectively. For our analysis, we have selected $n=3$ as an appropriate number of Gaussian components necessary to accurately model the amplitude distribution of the received signal under pheromone stimulation conditions (where $j=1$). By fitting this distribution to our data, we have derived the parameters outlined in Table \ref{GMM params 1}.

\setlength{\tabcolsep}{2pt}
\begin{table}[!h]
  \caption{Gaussian Mixture Model with three components}
  \label{GMM params 1}
  \centering
  \begin{tabular}{|c|c|c|c|c|c|c|c|c|}
    \hline
    $\pi_{11}$&$\mu_{11}$&$\sigma_{11}$&$\pi_{12}$&$\mu_{12}$&$\sigma_{12}$&$\pi_{13}$&$\mu_{13}$&$\sigma_{13}$\\
    \hline
0.2737&0.3307&0.0065&0.0439&0.7878&0.0152&0.6824&0.2323&0.0015\\
\hline
   \end{tabular}
\end{table}

The  "leaky" \textit{FUS1} expression (j=0) can also be described by a GMM, with two components. The corresponding parameters are mentioned in Table \ref{GMM params 0}.

\setlength{\tabcolsep}{2pt}
\begin{table}[!h]
  \caption{Gaussian Mixture Model with two components}
  \label{GMM params 0}
  \centering
  \begin{tabular}{|c|c|c|c|c|c|}
   \hline
$\pi_{01}$&$\mu_{01}$&$\sigma_{01}$&$\pi_{02}$&$\mu_{02}$&$\sigma_{02}$\\
\hline
    0.3816&0.0482&0.00074&0.6185&0.269&0.016\\
    \hline
\end{tabular}
\end{table}

\begin{figure}[!hbt]
\centering
\includegraphics[width=3in]{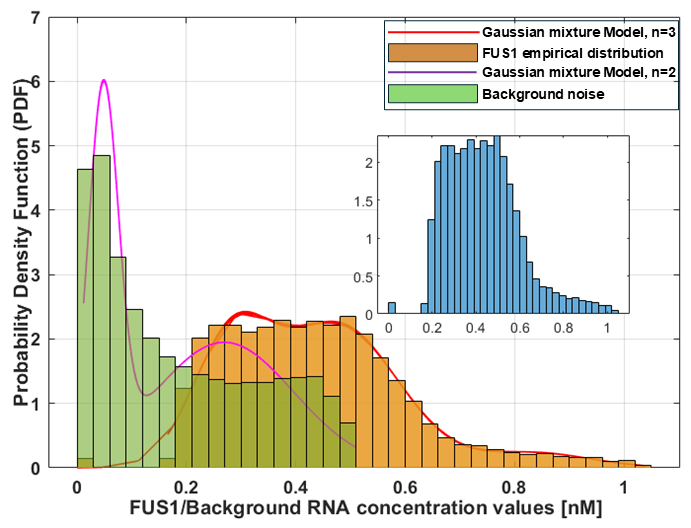}
\caption{Empirical distributions fitted using Gaussian models. The inset provides a clearer representation of the empirical distribution for the up-regulation values of \textit{FUS1}, which demonstrates a long tail.} 
\label{distributions}
\end{figure}

To implement Maximum Likelihood Detection effectively, it is essential to characterize the probabilistic distribution of the observed gene expression levels, whether the Rx is stimulated by  $\alpha$-factor molecules or not. In Figure \ref{distributions}, we present the empirical distributions of the two datasets. Additionally, we illustrate $g_1(x|\mu_1,\sigma^2_1)$ and $g_0(x|\mu_0,\sigma^2_0)$,  incorporating the parameters defined in Tables \ref{GMM params 1} and \ref{GMM params 0}. 

For clarity, we denote the instances of transmission and non-transmission as $s_1$ and $s_0$, respectively. The Maximum Likelihood Detection (MLD) criterion employs the Likelihood Ratio Test (LRT), under the assumption that $s_1$ and $s_0$ are equally probable, specifically $P(s_1)=P(s_0)=\frac{1}{2}$ \cite{kay1998fundamentals}.  This leads us to the following mathematical formulation:

\begin{equation} \label{MLD}
LRT=\frac{g_1(x|\mu_1,\sigma^2_1)}{g_0(x|\mu_0,\sigma^2_0)} \gtreqless  1
\end{equation}

In the case where $LRT \geq 1$, the up-regulation of (\textit{FUS1}) indicates transmission, while values below this threshold correspond to basal noise. To determine the threshold value, we identify the intersection point where the two likelihoods are equal. The resulting equation involves the sum of Gaussian terms, making its solution mathematically complex. Therefore, we employ computational methods using MATLAB. We found the threshold to be approximately 0.25 nM, meaning that any expression levels above this theoretical limit are interpreted as $\alpha$-factor transmission.

\subsection{Error Probability}
As part of our detection methodology, we analyze the probability of erroneous reception, denoted as $P(e|s)$, in the context of transmitting a signal $s$. Since we are dealing with binary transmission ("0" and "1"), we differentiate between two scenarios: the detection of a transmission ("1") when no transmission has actually occurred ($P(e|0)$), and the failure to detect a transmission ("0") when transmission has occurred ($P(e|1)$).

After determining the decision threshold, represented as $T$, we proceed to formulate the probabilities for these two distinct cases, as outlined in equations \eqref{Pe0} and \eqref{Pe1}. To compute each of these probabilities, we integrate the corresponding probability density functions (pdf) within the appropriate bounds that define the decision regions for each scenario. To accurately establish these bounds, we define $x_1$ as the minimum expression level of \textit{FUS1} under $\alpha$-factor stimulation, and  $x_2$  as the maximum expression level of \textit{FUS1} in the absence of pheromone stimulation.

\begin{equation}\label{Pe0}
\begin{aligned}
&P(e|0)= \int_{T}^{x_2} g_0(x|\mu_0,\sigma^2_0) \,dx = \int_{T}^{\infty} g_0(x|\mu_0,\sigma^2_0) \,dx - \\& \int_{x_2}^{\infty} g_0(x|\mu_0,\sigma^2_0) \,dx = \int_{T}^{\infty} (\pi_{01}\frac{1}{\sqrt{2\pi}\sigma_{01}}e^{-\frac{1}{2}(\frac{x-\mu_{01}}{\sigma_{01}})^2} + \\ &\pi_{02}\frac{1}{\sqrt{2\pi}\sigma_{02}}e^{-\frac{1}{2}(\frac{x-\mu_{02}}{\sigma_{02}})^2}) \,dx - \int_{x_2}^{\infty} (\pi_{01}\frac{1}{\sqrt{2\pi}\sigma_{01}}e^{-\frac{1}{2}(\frac{x-\mu_{01}}{\sigma_{01}})^2} + \\ &\pi_{02}\frac{1}{\sqrt{2\pi}\sigma_{02}}e^{-\frac{1}{2}(\frac{x-\mu_{02}}{\sigma_{02}})^2}) \,dx
\end{aligned}
\end{equation}

We group the terms multiplied by the coefficients $\pi_{01}$ and $\pi_{02}$ respectively, and we make use of the Q function formula: $Q(x) = \frac{1}{\sqrt{2\pi}}\int_{x}^{\infty} e^{-\frac{t^2}{2}} \,dt$. After substituting $t=\frac{x-\mu}{\sigma}$ and modifying the integrals in \eqref{Pe0} accordingly, we result in a sum of Q functions of the following form:

\begin{equation}\label{Pe0_2}
\begin{aligned}
&P(e|0) = \pi_{01}\left(Q\left(\frac{T-\mu_{01}}{\sigma_{01}}\right)- Q\left(\frac{x_2-\mu_{01}}{\sigma_{01}}\right)\right) + \\& \pi_{02}\left(Q\left(\frac{T-\mu_{02}}{\sigma_{02}}\right)-Q\left(\frac{x_2-\mu_{02}}{\sigma_{02}}\right)\right)&
\end{aligned}
\end{equation}

Similarly, when a "0" is detected while transmitting a "1", the equation above can be modified to include the corresponding integration limits, as shown in \eqref{Pe1}.

\begin{equation}\label{Pe1}
\begin{aligned}
&P(e|1) = \int_{x_1}^{T} g_1(x|\mu_1,\sigma^2_1) \,dx =  \int_{-\infty}^{T} g_1(x|\mu_1,\sigma^2_1) \,dx - \\& \int_{-\infty}^{x_1} g_1(x|\mu_1,\sigma^2_1) \,dx = \int_{-\infty}^{T} (\pi_{11}\frac{1}{\sqrt{2\pi}\sigma_{11}}e^{-\frac{1}{2}(\frac{x-\mu_{11}}{\sigma_{11}})^2} + \\ &\pi_{12}\frac{1}{\sqrt{2\pi}\sigma_{12}}e^{-\frac{1}{2}(\frac{x-\mu_{12}}{\sigma_{12}})^2} + \pi_{13}\frac{1}{\sqrt{2\pi}\sigma_{13}}e^{-\frac{1}{2}(\frac{x-\mu_{13}}{\sigma_{13}})^2}) \,dx - \\& \int_{-\infty}^{x_1} (\pi_{11}\frac{1}{\sqrt{2\pi}\sigma_{11}}e^{-\frac{1}{2}(\frac{x-\mu_{11}}{\sigma_{11}})^2} + \\ &\pi_{12}\frac{1}{\sqrt{2\pi}\sigma_{12}}e^{-\frac{1}{2}(\frac{x-\mu_{12}}{\sigma_{12}})^2} + \pi_{13}\frac{1}{\sqrt{2\pi}\sigma_{13}}e^{-\frac{1}{2}(\frac{x-\mu_{13}}{\sigma_{13}})^2}) \,dx
\end{aligned}
\end{equation}

Which can also be reduced in the following:

\begin{equation}\label{Pe1_2}
\begin{aligned}
&P(e|1) = \pi_{11}\left(Q\left(\frac{\mu_{11}-T}{\sigma_{11}}\right)-Q\left(\frac{\mu_{11}-x_1}{\sigma_{11}}\right)\right) + \\& \pi_{12}\left(Q\left(\frac{\mu_{12}-T}{\sigma_{12}}\right)-Q\left(\frac{\mu_{12}-x_1}{\sigma_{12}}\right)\right) + \\& \pi_{13}\left(Q\left(\frac{\mu_{13}-T}{\sigma_{13}}\right)- Q\left(\frac{\mu_{13}-x_1}{\sigma_{13}}\right)\right)&
\end{aligned}
\end{equation}

To determine the total Probability of Error $Pe$, we must consider both $P(e|0)$ and $P(e|1)$  while accounting for their respective prior probabilities.

\begin{equation} \label{Pe}
P_e = P(1)P(e|1) + P(0)P(e|0)
\end{equation}

\begin{figure}[!hbt]
\centering
\includegraphics[width=3in]{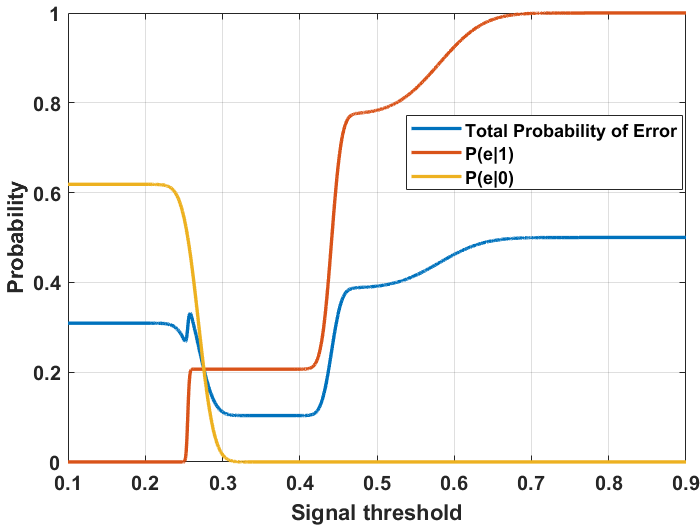}
\caption{Probability of false alarm (yellow), hit miss (orange) and total erroneous reception (blue).}
\label{Pe plots}
\end{figure}

Assuming equal prior probabilities, $P(1)=P(0)=\frac{1}{2}$, Figure \ref{Pe plots} illustrates the variability of  $Pe$, $P(e|1)$  (hit miss), and  $P(e|0)$ (false alarm). By adjusting the threshold, we can minimize $P_e$ and optimize system performance. However, it is observed that the minimum achievable $P_e$ is around 10\%, which is relatively high.

Biological systems employed for MC are inherently limited in their BER performance due to the noisy nature of biological processes. Background noise, arising from "leaky" gene expression and inherent noise in processes like transcriptional binding and molecular diffusion, significantly impacts performance.

To effectively mitigate noise effects, careful engineering of the MATa Rx cell is crucial. This involves strategies to suppress background noise and enhance the cell's ability to upregulate \textit{FUS1} in response to external stimulation with $\alpha$-factor.

\subsection{Sequence Detection}

Building upon the detection framework for single symbols and considering On-Off Keying (OOK) modulation with one bit per symbol, we extend our analysis to develop a method for detecting sequences of symbols. In this biological system, molecular diffusion introduces memory into the channel, causing ISI. As a result, the received symbols cannot be considered independent.

\begin{figure}[!hbt]
\centering
\includegraphics[width=3in]{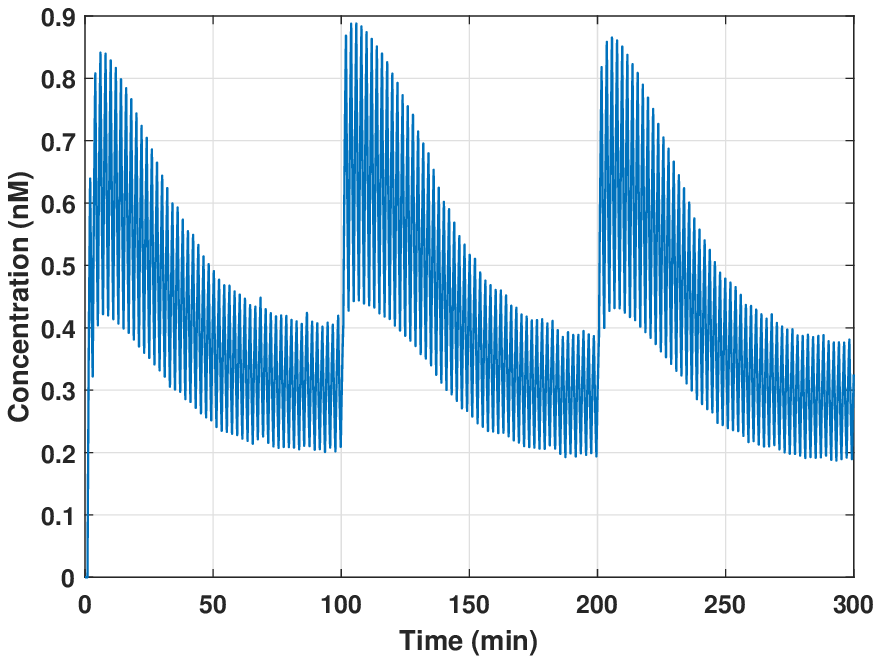}
\caption{Received symbols in the form of \textit{FUS1} output patterns for $k=3$.}
\label{FUS1 pulses}
\end{figure}

To address this challenge, we propose a theoretical framework for sequence detection using the Maximum Likelihood Sequence Detection (MLSD) approach \cite{kay1998fundamentals}. This framework quantifies the likelihood of a transmitted sequence based on the Euclidean Distance between the transmitted and received sequences. Minimizing this distance essentially aligns the received noisy sequence with the most probable transmitted sequence. The mathematical formulation of this problem is as follows:

\begin {equation}\label{Eucl}
\underset{(s_j^{(1)},s_j^{(2)},s_j^{(3)},...,s_j^{(k)})}{\text{arg \ min}}\,\,\,\, \sum_{i=1}^{k} ||s_j^{(i)}-r_j^{(i)}||^2
\end{equation}

where k represents the number of transmitted signals (symbols), and j denotes the sequence index. In this study, we consider symbol sequences consisting of three symbols, resulting in $k=3$ and $1\leq j \leq8$. Repetitive pheromone stimulation at the Rx generates distinct \textit{FUS1} upregulation patterns. These readouts serve as the received symbols that undergo detection. For $k=3$ $\alpha$-factor stimulations, the received \textit{FUS1} pulses corresponding to these symbols are depicted in Figure \ref{FUS1 pulses}.

To evaluate the performance of the proposed framework, we analyze the achievable symbol error rates while considering the two primary noise sources discussed in Section \ref{Gene Noise}. To this end, all possible transmitted symbol sequences, ranging from 000 to 111, were evaluated. For each transmitted sequence, the mathematical model was employed to compute the Rx response, incorporating both ISI and noise effects.

The biological yeast Rx upregulates the \textit{FUS1}  gene following each stimulation. Consequently, the received symbols are characterized by the time evolution dynamics imposed by the channel model and are represented as time vectors. To account for this characteristic, we modify Equation \ref{Eucl} as follows:

 \begin {equation}\label{Eucl_time}
\underset{(s_j^{(1)},s_j^{(2)},s_j^{(3)},...,s_j^{(n)})}{\text{arg \ min}}\,\,\,\, \sum_{i=1}^{n} ||s_j^{(i)}-r_j^{(i)}||^2
\end{equation}

where n represents the total number of time points considered in the computations. The input sequence that minimizes Equation \ref{Eucl_time} is deemed the transmitted sequence.

To characterize the error occurrence rates for a transmitted sequence of three symbols, we compute the Symbol Error Rate (SER) for each of the eight possible sequences $1\leq j \leq8$ as follows:

 \begin {equation}\label{SER}
SER = \frac{n_f}{n_{total}}
\end{equation}

The number of falsely detected symbols is denoted as $n_f$, while the total number of received symbols is represented by $n_{total}$.
To assess the impact of noisy dynamics on the received symbol set, the Symbol Error Rate (SER) is quantified using Equations \ref{Eucl_time} and \ref{SER} with 500 iterations.

\begin{figure}[!hbt]
\centering
\includegraphics[width=3in]{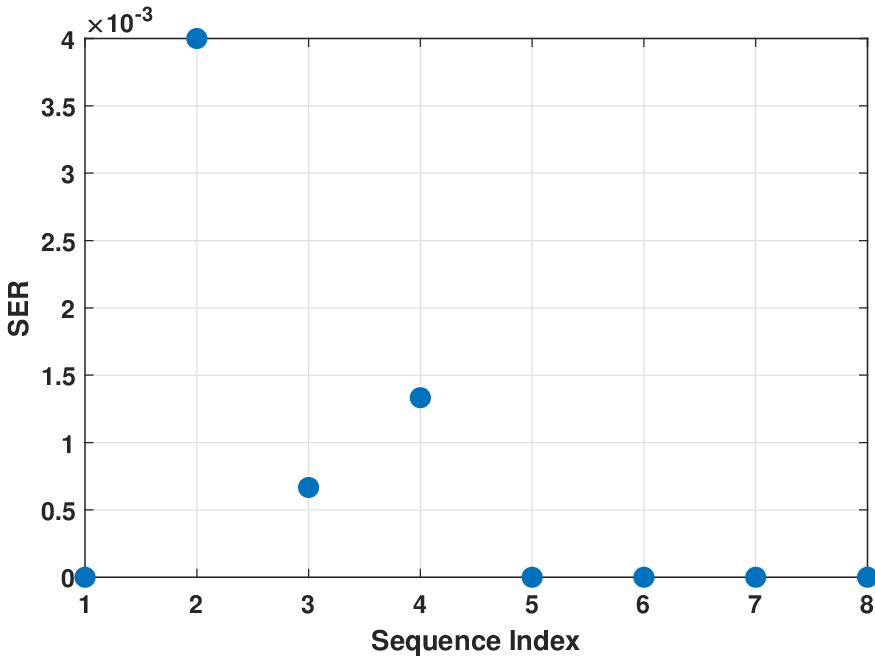}
\caption{Symbol Error Rates for Each of the Eight Possible Symbol Sequences with $k=3$ Transmitted Symbols}
\label{SER All Sequences}
\end{figure}

Across all symbol combinations, the computed SER demonstrates that the MLSD framework exhibits high detection reliability, achieving minimal error rates. However, the computational complexity of the proposed framework may become a bottleneck when applied to longer symbol sequences. The average SER for all possible sequences, considering $k=3$, is presented in Figure \ref{SER All Sequences}. 

Additional simulations were performed to examine the impact of varying $\alpha$-factor concentrations on the maximum Symbol Error Rate (SER) achieved across eight distinct transmitted symbol sequences. The results, presented in Table \ref{sims} in percentage form, demonstrate a consistent increase in the maximum SER for all tested $\alpha$-factor concentrations.

\setlength{\tabcolsep}{2pt}
\begin{table}[!h]
  \caption{Maximum SER increase for different pheromone concentration values}
  \label{sims}
  \centering
  \begin{tabular}{|c|c|c|}
   \hline
$\alpha$=20$\mu$M&$\alpha$=60$\mu$M&$\alpha$=80$\mu$M\\
\hline
    1150\%&1575\%&1650\%\\
    \hline
\end{tabular}
\end{table}

This upward trend in SER provides strong evidence of the presence of non-linear dynamics, which become particularly pronounced when the $\alpha$-factor concentration exceeds 10$\mu$M. Furthermore, the data in Table \ref{sims} reveal that the SER values are closely clustered, suggesting the onset of saturation effects. This observation aligns with findings from our previous work \cite{10429943} and the research conducted in \cite{Kofahl2004ModellingTD}, where a pheromonal concentration of 10$\mu$M was identified as the threshold beyond which saturation phenomena begin to manifest. These results underscore the importance of considering non-linear dynamics and saturation effects in the analysis of SER, particularly at higher $\alpha$-factor concentrations.

Compared to symbol-by-symbol detection, sequence detection has demonstrated significantly lower error rates—approximately two orders of magnitude improvement—when utilizing MLSD. This enhancement stems from the inherent limitations of symbol-by-symbol detection. In this approach, biological background gene expression, occurring even in the absence of physical stimulation, coupled with the stochastic nature of transcriptional binding and unbinding events (Poissonian distribution), increases the probability of erroneously interpreting a "0" as a "1. Conversely, sequence detection leverages Euclidean distances to directly compare signals, thereby mitigating the influence of noise more effectively. However, this improved accuracy comes at the cost of increased computational complexity.
 
\section{Conclusion}\label{Conclusion}

In this work we investigate Maximum Likelihood Detection (MLD) within a yeast-based experimental platform, focusing on the impact of noise. Transcription factor binding is identified as the primary source of noise in pheromone-stimulated signaling pathways, and this observation is integrated into a validated channel model.

Noise is modeled as a birth-death process, represented by Poisson arrivals and departures, accurately reflecting experimental observations. Consequently,  \textit{FUS1} gene expression exhibits stochastic variations, best described by Mixed Gaussian processes.

We characterize MLD in terms of Bit Error Rates, finding that optimizing the signal threshold is crucial for minimizing the Probability of Error due to the unique nature of gene expression levels. The estimated lower error bound, approximately 10\%, is primarily attributed to the inherently noisy environment of yeast pheromone communication.

Conversely, employing pheromonal Rx re-stimulation and utilizing MLSD significantly improves performance, achieving an upper error rate bound of approximately 0.4\%. However, MLSD increases computational complexity, potentially posing challenges for longer symbol sequences.

Future work will delve deeper into significant noise sources, particularly at the sender side. Experimental studies will validate the incorporated noise models and demonstrate the effectiveness of the proposed detection policy in terms of achievable bit rates. The impact of nonlinear effects on system performance, arising from the overstimulation of the Rx, will be thoroughly investigated to establish an optimal trade-off between the transmitted concentration of the signaling factor and the accuracy of detection. This analysis will provide valuable insights into balancing the intensity of the transmitted signal with the receiver's operational limits, ensuring robust and reliable detection accuracy while avoiding the adverse consequences of overstimulation. Such a trade-off is essential for optimizing system performance in practical applications where signal integrity and detection precision are paramount.

\section{Acknowledgments}
This research has received support from the project BRIDGE2HORIZON/0823D/004 (YeasPSense) which is implemented under the Cohesion Policy Funds “THALIA 2021-2027” with EU co-funding.



\bibliography{Refs}{}
\bibliographystyle{IEEEtran}

\vfill

\end{document}